# Ethylene Carbonate Adsorption and Decomposition on Pristine and Defective ZnO (10$\bar{1}$0) Surface: A First-Principles Study

Ganes Shukri[1,2],\*, Bernardus Rendy[1], Adhitya Gandaryus Saputro[1,2],\*, Febriyanti Veren Panjaitan[1], Poetri Sonya Tarabunga[1], Mohammad Kemal Agusta[1,2], Nadhratun Naim Mobarak[3], and Hermawan Kresno Dipojono[1,2]

[1]Advanced Functional Materials Research Group, Faculty of Industrial Technology, Bandung Institute of Technology, Bandung 40132, West Java, Indonesia

[2]Research Center for Nanoscience and Nanotechnology (RCNN), Bandung Institute of Technology, Bandung 40132, West Java, Indonesia

[3]Department of Chemical Sciences, Faculty of Science and Technology, Universiti Kebangsaan Malaysia, Bangi 43600, Selangor, Malaysia

**Abstract**

Fundamental understanding of the reactivity between coating material of Li-ion battery cathode and electrolyte is important in order to obtain suitable coating candidates. Herein, we study ethylene carbonate (EC) adsorption and decomposition reactions on pristine, O vacancy- and Zn vacancy-defected ZnO (10$\bar{1}$0) by means of first-principles density functional theory (DFT) calculations. Possible decomposition pathways via H-abstraction and EC ring-opening reaction that leads to the generation of $CO_2$ and $C_2H_4$ gases are studied from the thermodynamic and kinetic aspects. Firstly, we find that molecular EC preferably adsorbs on both pristine and defective ZnO ($10\bar{1}0$) via the bonding between its carbonyl oxygen ($O_C$) and surface Zn. Secondly, subsequent decomposition reactions show large tendency of EC to decompose on both pristine and defective ZnO ($10\bar{1}0$). This tendency is indicated by the large thermodynamic driving forces to decompose EC that range from -1.5 eV to -2.5 eV on both pristine and defective ZnO ($10\bar{1}0$) (calculated with respect to EC gas phase). The large tendency of EC to decompose, however, is hindered by the high activation barriers of the EC decomposition, shown by the lowest activation barrier of 0.96 eV on Zn vacancy defected ZnO ($10\bar{1}0$). Our results thus indicate that EC decomposition on ZnO ($10\bar{1}0$) is mainly hindered due to its slow rate of decomposition instead of the thermodynamic factors.

1. Introduction

High energy cathode for Li-ion battery (LIB) such as Ni-rich NCM, NCA, and Li-rich layered oxides are in high demand for electric vehicle and grid energy storage applications. To ensure their energy capacity can be utilized up to near their theoretical values, it is required to go beyond the current typical commercially LIB operating limit, i.e., State of Charge (SOC) > 50% [1,2]. However, the required condition is hindered to attain mainly due to rapid loss of cathode energy capacity in particular when the LIB is cycled to high voltage, i.e., > ~ 4.4 $V_{Li}$ [3–5]. One of the main origins of the capacity loss of layered cathode is structural deterioration due to irreversible phase transformation from layered to disordered spinel/rock-salt. This phase transformation is commonly initiated from the cathode surface and subsequently propagates into the bulk, leading to a more densified structure that may impede ionic mobility inside the cathode [6–8]. Previous studies have observed that the phase transformation of the cathode is closely related to the electrochemical/chemical oxidation of the liquid electrolyte catalyzed by the cathode surface. For instance, by using On-Line Electrochemical Mass Spectroscopy (OEMS), Jung *et al* proposed a mechanism in which surface lattice oxygen of layered oxide cathodes (viz., NMC111, NMC622, NMC811 and LNMO) induces liquid carbonate decomposition into carbon- and oxygen-containing gases (e.g., CO, $CO_2$, $C_2H_4$) [9]. The then released lattice oxygen creates a vacancy defect that may accelerate metal cation migration to the Li layer, eventually leads to the detrimental rock-salt phase formation on the cathode [7,10,11]. Further, the products of cathode-electrolyte interaction are found to also trigger rapid growth of the cathode-electrolyte interface (CEI) layer [12–14]. The uncontrolled side reaction of electrolyte decomposition on the CEI layer combined with irreversible phase transformation of the cathode can eventually degrade the overall electrochemical performance of the LIB. In addition, there is also safety concern with regard to the built-up pressure inside the LIB cell due to the residual gas produced by the electrolyte decomposition.

Depositing a protective coating layer on the LIB cathode surface is a promising way to enhance the cathode stability at a high operating voltage. Various compounds such as metal oxides [15–17], phosphates [18], and fluorides [19] have been reported as potential protective coating materials for LIB cathode. Several mechanisms have been proposed in order to explain the enhanced stability of the coated cathode: (1) the coating scavenges HF molecule, thus decreasing local acidity at the cathode-electrolyte interface area [20], (2) the coating shields the active cathode materials from deleterious (electro)chemical decomposition reactions of the electrolyte [21,22] and (3) the coating mitigates irreversible structural deformation on the cathode due to surface oxygen loss, transition metal dissolution and surface cation migration [23–25]. Based on the aforementioned proposed mechanisms, several studies have even attempted to systematically screen candidates of coating materials with robust thermodynamic stability against HF formation and LiF precipitation [26]. There were other studies as well that suggest the Li and defect diffusion rate as an important criterion for coating materials [27,28]. Those previous studies indeed have given valuable insights regarding some important physical and chemical properties that should be possessed by a good coating material.

While studies related to HF-scavenging and ionic diffusion properties on candidates of coating material have been intensively carried out, studies related to how the carbonate electrolyte molecule interacts with the coating layer are relatively still scarce. This is mainly due to the challenging task on probing details of the reaction between the protective coating later with the carbonate molecule. While previous studies have shown that carbonate molecules (e.g., EC, DMC) can react with the cathode surface to produce gaseous compounds [29–31], the detailed reaction of the carbonates with the coating layer is not yet studied in detail. This study is important in order to assess the ability of the coating layer to also decrease the decomposition of carbonates. Understanding the thermodynamic and kinetic of carbonate molecule decomposition on the coating layer surface is important to obtain optimum design rules of the protective coating layer.

In this work, we investigate ethylene carbonate (EC) decomposition on pristine and defective ZnO ($10\bar{1}0$) surfaces. ZnO is chosen as a model for the protective coating layer because previous studies have shown the increased electrochemical performance of LIB composed of ZnO-coated cathodes [31–33]. We start by evaluating all thermodynamically stable adsorption configurations on both pristine and defective ZnO ($10\bar{1}0$) surfaces. Two surface point defects are considered, i.e., the O and Zn vacancies ($O_{vac}$ and $Zn_{vac}$). Our consideration to include point defects is based on previous reports that most protective coating layer is in the amorphous form [28,34,35]. Thus, the inclusion of point defect in our crystalline ZnO model is an attempt to include the under-coordinated local Zn-O bonds that commonly exist in an amorphous ZnO (as well as any other amorphous oxides). We then investigate several possible EC decomposition reactions that lead to the generation of carbon-containing gases, viz., $CO_2$, $CH_2CHO$ and $C_2H_4$. These three carbon-containing gases are chosen because they are frequently observed in various *in-situ* and post-mortem analyses of cycled LIB [9,29,30,36–38]. To conclude the study, we compare the obtained thermodynamic and kinetic properties of EC decomposition on pristine and defective ZnO ($10\bar{1}0$) of this work with previous EC decomposition on several LIB cathodes.

## 2. Computational Details

We carried out spin-polarized density functional theory (DFT) calculations within the projector augmented wave (PAW) formalism as implemented in the Quantum ESPRESSO 6.1 package [39]. Generalized gradient approximation of Perdew-Burke-Ernzerhoff (GGA-PBE) was used to describe the exchange interactions and electronic correlations [40]. Total electronic energy convergence was achieved by using cutoff energy of 42 Ry for the expansion of Kohn-Sham wavefunctions. Brillouin zone integration for all geometry and electronic structure calculations were performed only at the gamma point considering the large lateral size of the slab used in the present study. We used Hubbard-$U$ correction of 7.5 eV for Zn 3$d$ states following previous study of reduced ZnO with oxygen vacancy [41]. We also considered the dispersion effect for all calculations using semiempirical correction proposed by Grimme (DFT-D2) [42]. 15 Å of vacuum region was added to the normal direction of the surface to prevent any spurious interaction between the periodically repeated images of the slab. We used a slab model with eight Zn-O layers with a 4 x 4 lateral size. To model the defective ZnO (10$\bar{1}$0) surface, we considered two types of point defect, viz., O vacancy ($O_{vac}$) and Zn vacancy ($Zn_{vac}$). The two defects were introduced on the topmost layer of the ZnO (10$\bar{1}$0) surface. Both the pristine and defective ZnO (10$\bar{1}$0) slab models are shown in **Fig 1a-g**. Note that all geometry optimizations were done by relaxing all atoms of EC molecule and the two topmost Zn-O layers while the rest were fixed to their optimized bulk positions.

The adsorption energy ($E_{ad}$) of an EC molecule on ZnO (10$\bar{1}$0) surface is defined as:

$$E_{ad} = E_{sys} - (E_{surf} + E_{mol}), \qquad (1)$$

where $E_{sys}$ corresponds to the total energy of the adsorption system, $E_{surf}$ corresponds to the total energy of the ZnO (10$\bar{1}$0) slab and $E_{mol}$ corresponds to the total energy of an isolated EC molecule. The activation energy for EC decomposition reaction ($E_{act}$) was calculated by using the following formula:

$$E_{act} = E_{TS} - E_{IS}, \qquad (2)$$

where IS and TS correspond to the initial and transition configuration of the decomposition reaction, respectively. The structure of the TSs was calculated by using the climbing image nudged elastic band method (CI-NEB) with four intermediate images. We would like to note that all EC adsorption and decomposition reactions studied in the present work were excluded from any external potential. Also, we did not include the presence of Li in our ZnO slab model. Thus, our present work can be thought as a model of EC-ZnO interaction that occur before any operating condition of the LIB is carried out.

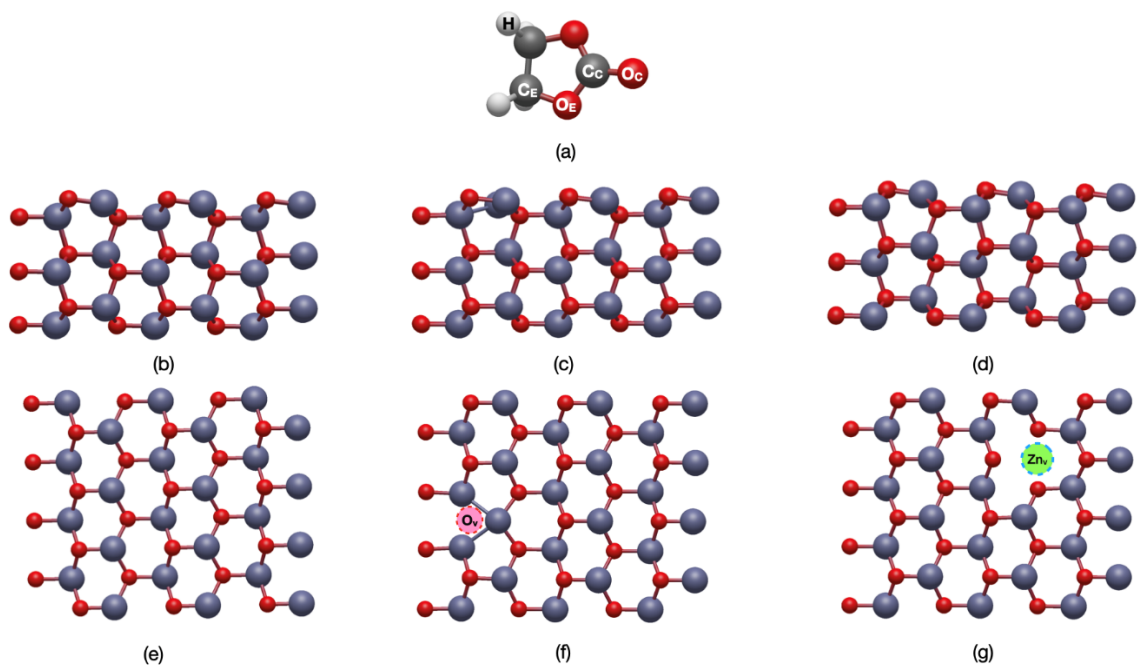

**Figure 1.** Structure of ethylene carbonate (a). Side view & top view of stoichiometric ZnO ($10\bar{1}0$) surface (b) & (e), surface with an O vacancy (c) & (f), and surface with a Zn vacancy (d) & (g). The location of surface O vacancy and Zn vacancy are labeled by $O_{vac}$ and $Zn_{vac}$.

## 3. Results and Discussions

### 3.1. EC Molecular Adsorption on Pristine and Defective ZnO ($10\bar{1}0$)

We started by calculating adsorption geometry and energy of a single EC molecule on both pristine and defective ZnO ($10\bar{1}0$) surface. Upon careful investigation, we found that the most stable EC adsorption configuration on pristine ZnO ($10\bar{1}0$) is facilitated by its $O_C$ binds to the surface Zn ($Zn_{surf}$) with adsorption energy of -1.28 eV (**Fig 2a**). This adsorption configuration is also observed on other layered oxide cathodes, e.g., NCM, LCO and LNO [43,44]. The $O_C$-$Zn_{surf}$ bonding can be further identified by the hybridization between the highest occupied molecular orbital (HOMO) of EC (composed of some $O_C$ 2$p$ states) with the initially empty $Zn_{surf}$ 3$d$ states as shown by the projected electronic density of states (PDOS) (**Fig. 3a**). Further, Lowdin charge analysis suggests that approximately 0.27$e$ are transferred from the EC to the ZnO upon adsorption. Previous studies also observed another possible adsorption configuration of EC on metal oxide surface where, along with $O_C$-$TM_{surf}$ (surface transition metal) bonding, the $C_C$ also bonds to the surface O ($O_{surf}$) of the metal oxide [43,45]. Indeed, we found similar configuration adsorption is also thermodynamically favorable with -0.84 eV adsorption energy (**SI Fig S1d**). The identification of these adsorption configurations is important to further study the EC decompositions as we will discuss in the next chapter.

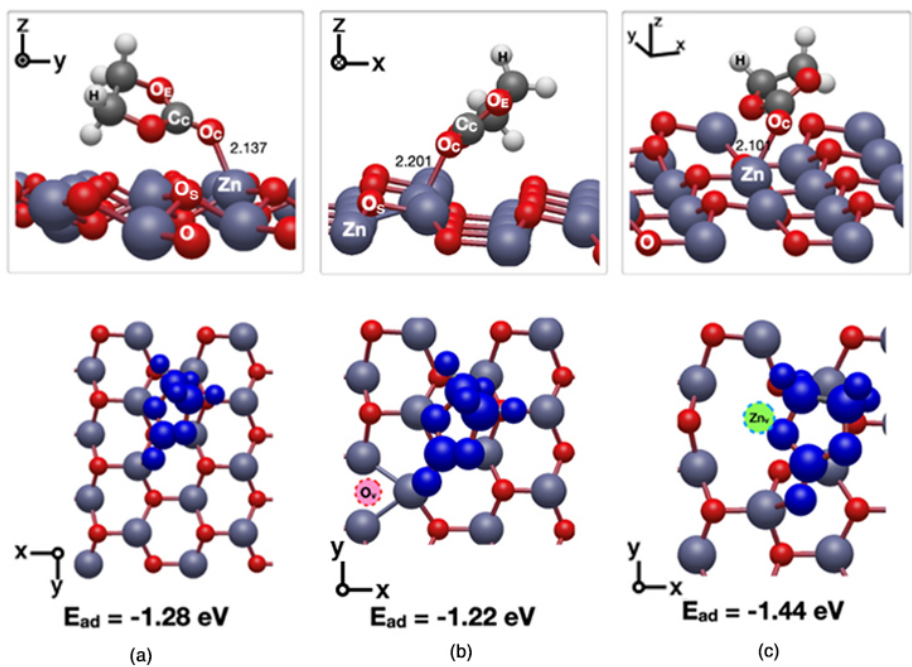

**Figure 2.** The side view and top view of the most stable EC adsorption structures on the (a) stoichiometric ZnO (10$\bar{1}$0) surface, (b) ZnO (10$\bar{1}$0) surface with an O vacancy, and (c) ZnO (10$\bar{1}$0) surface with a Zn vacancy. The collection of blue atoms in the second row (top view) indicates the adsorbed EC molecule.

For the case of EC adsorption on ZnO (10$\bar{1}$0) with O vacancy (O$_{vac}$) we found that the most stable adsorption structure is similar to the stoichiometric case where O$_C$ bonds with the under-coordinated three-fold Zn$_{surf}$ (**Fig. 2b**). The adsorption energy is calculated to be -1.22 eV. PDOS plot again indicates the hybridization between the HOMO of EC with the Zn$_{surf}$ 3$d$ states (**Fig. 3b**). We also note that the 'ethylene group' of EC tends to be positioned further from the O$_{vac}$ site upon adsorption. This is so to minimize the repulsive interaction between the excess electrons residing in reduced Zn near the O$_{vac}$ with the ethylene group of EC [41,46–48].

Next, for the case of molecular EC adsorption near surface Zn vacancy (Zn$_{vac}$), we found similar most stable EC adsorption configuration with the previous two cases where O$_C$ bonds to the four-fold Zn$_{surf}$ located nearest from the Zn$_{vac}$ site (**Fig. 2c, Fig. 3c**). Here, EC adsorbs with a slight offset to the bridges between surface atoms. Under the absence of Zn atom for O$_E$-Zn bonding, EC tilts the same with O$_E$ closing to the Zn vacancy location. From other thermodynamically favorable adsorption configurations (**ESI Fig. S3**), those with O$_E$ close to Zn$_{vac}$ location have stronger adsorption except the second most stable configuration.

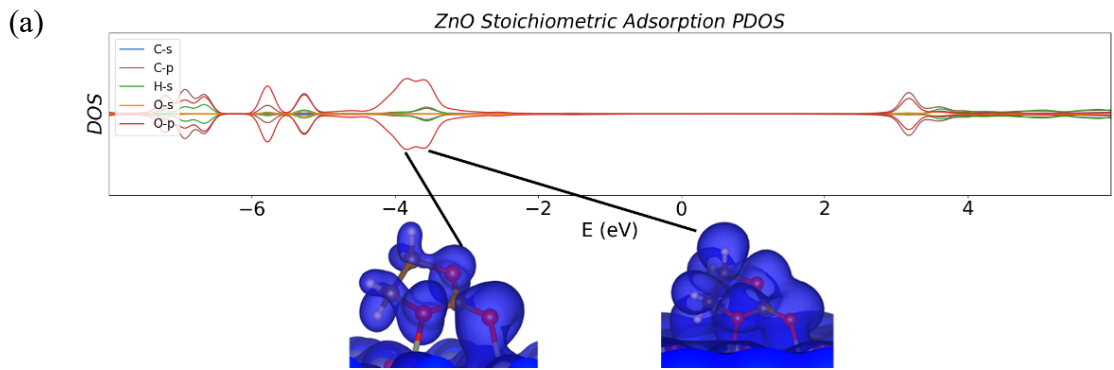

(a) ZnO Stoichiometric Adsorption PDOS

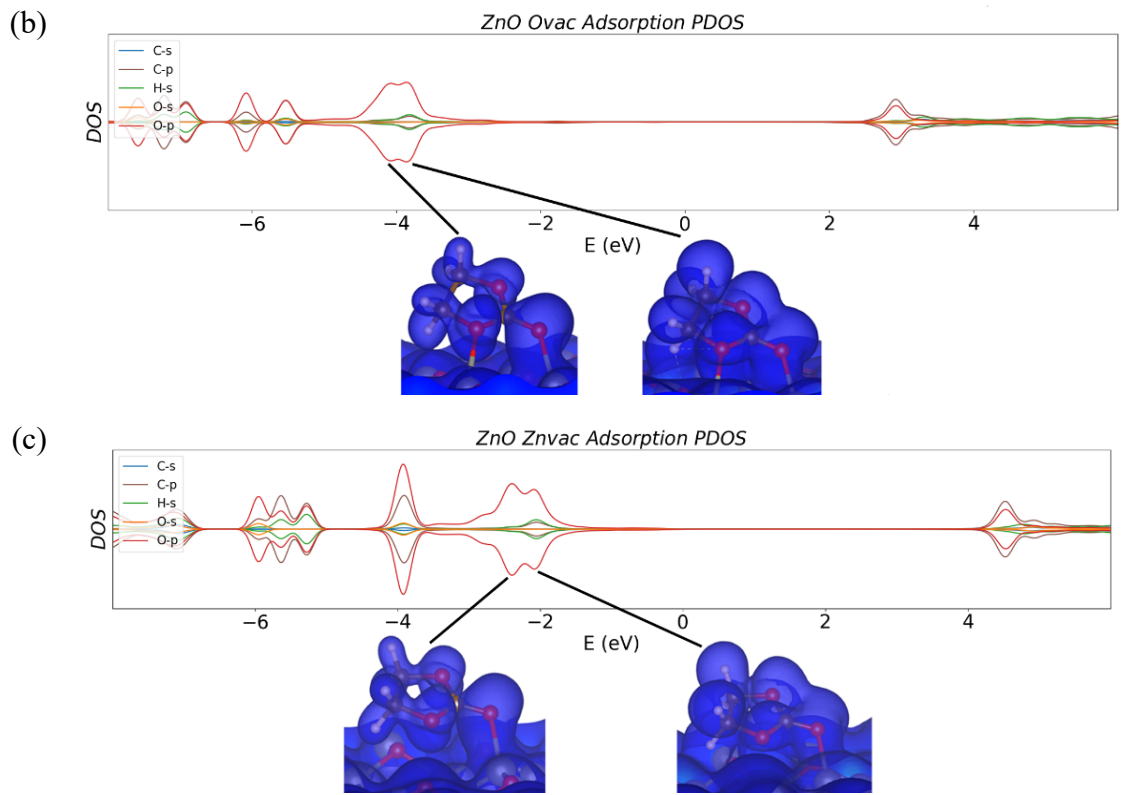

**Figure 3.** Projected density of states (PDOS) and the corresponding orbital hybridization (overlap) through partial charge density isosurface on the peaks near Fermi energy in the PDOS for EC adsorption on the ZnO (10$\bar{1}$0) (a) stoichiometric surface, (b) surface with an O$_{vac}$, and (c) surface with a Zn$_{vac}$. Details of the EC isolated molecular orbital and charge density projections of several lower EC-ZnO derived-states are shown in **ESI Fig. S4 a-d**.

## 3.2. EC Decomposition Paths on Pristine ZnO (10$\bar{1}$0)

Based on the thermodynamically most favorable EC adsorption configurations, we then investigated EC decomposition reactions on both pristine and defective ZnO (10$\bar{1}$0). For the case of stoichiometric surface, we investigated three reactions (denoted as reaction A, B and C in **Fig. 4**). Details of the decomposition and adsorption of the decomposed products are designated by the given numbers. Number '1', '2' and '3' indicate the $C_{E2}$-$O_{E2}$ ring-opening reaction, the $C_{E1}$-$O_{E1}$ ring-opening reaction and the $C_{E1}$-$H_1$ deprotonation of EC ethylene group, respectively. The '4' and '5' indicate $O_{E1}$ adsorption on the $Zn_{surf}$ and $C_C$ adsorption on the $O_{surf}$, respectively.

The '*' sign indicates species that is adsorbed on the ZnO surface. Note again that all three decomposition reactions started from the most stable adsorption configuration as previously explained in section **3.1**. Unless otherwise stated, the zero-energy reference for the subsequent EC decomposition reaction plots are the isolated gas phase EC and isolated ZnO (10$\bar{1}$0) slab.

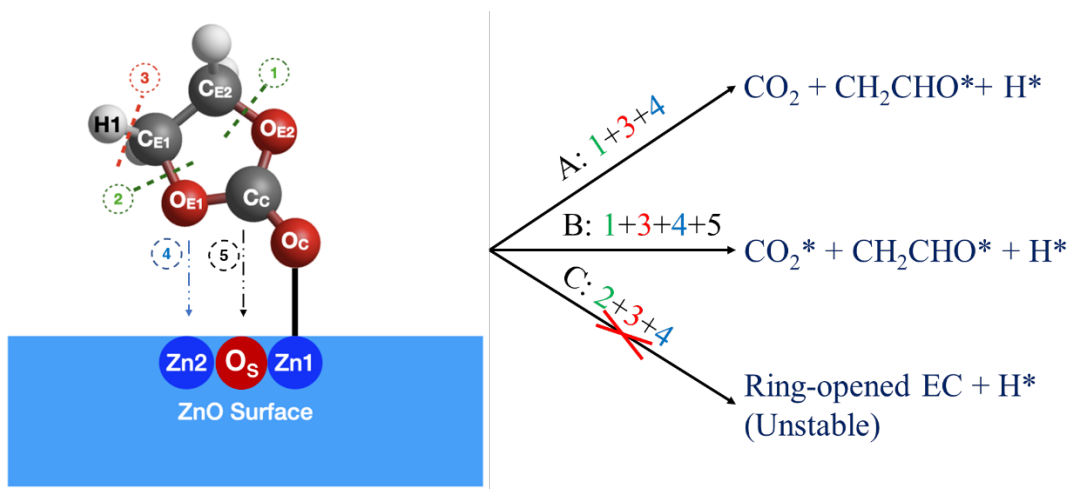

**Figure 4.** Schematic of initial decomposition reactions of EC on the stoichiometric ZnO (10$\bar{1}$0) surface.

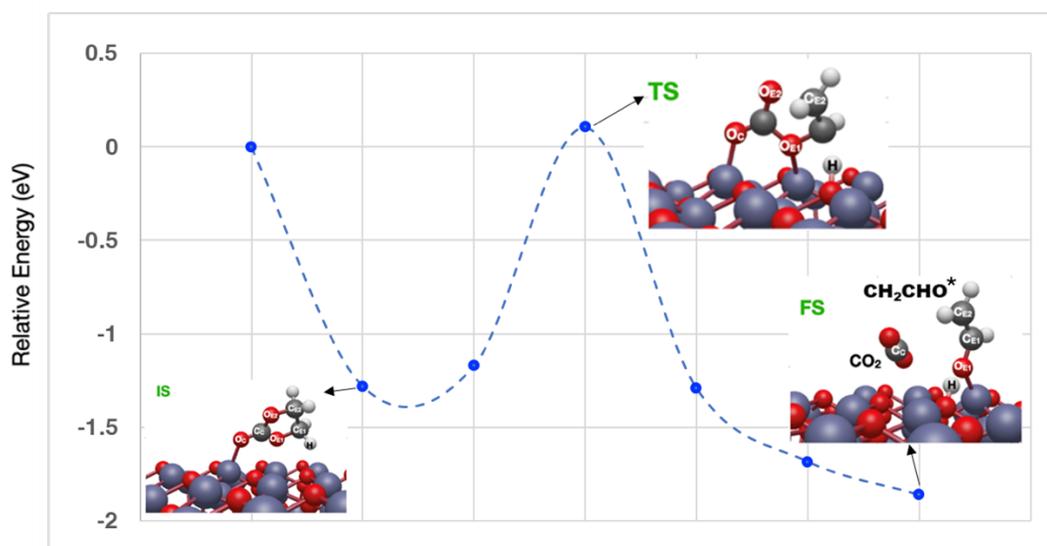

**Figure 5.** Potential energy curve for EC decomposition to $CO_2$, $CH_2CHO^*$, and $H^*$ (reaction A: 1+3+4) on the stoichiometric ZnO (10$\bar{1}$0) surface.

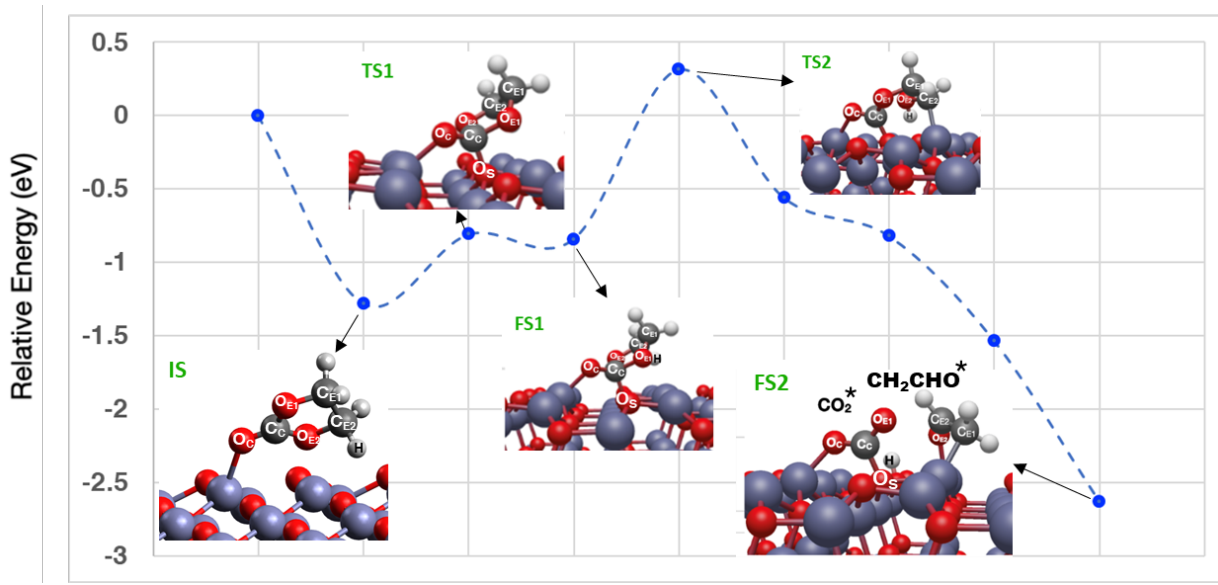

**Figure 6.** Potential energy curve for EC decomposition to $CO_2^*$, $CH_2CHO^*$, and $H^*$ (reaction B: 1+3+4+5) on the stoichiometric ZnO $(10\bar{1}0)$ surface.

Of the three reactions, reactions A and B are exothermic with overall reaction energies of -0.58 eV and -1.35 eV (with respect to EC adsorption state, i.e., 'IS'), respectively. While both reactions have similar EC decomposition products, notice that the transition and the final states configurations are different. EC decomposition via reaction A proceeds through simultaneous processes of $H_1$ abstraction and $C_{E2}$-$O_{E2}$ ring-opening reaction with 1.39 eV reaction barrier (**Fig. 5**). The end products are $CO_2$, $CH_2CHO^*$ (adsorbed on ZnO via $O_{E1}$-$Zn_{surf}$ bonding) and $H^*$-abstraction (eventually forming a surface hydroxyl). Reaction B proceeds through two steps (**Fig. 6**). The first step is $C_C$-$O_{surf}$ bonding with 0.44 eV reaction energy and 0.47 eV activation barrier (calculated with respect to the IS in **Fig. 6**). This step is similar to EC adsorption configuration found in NMC type cathode material [43]. The reaction then proceeds to the second step where EC decomposed into $CO_2^*$, $CH_2CHO^*$ (adsorbed on ZnO via $O_{E1}$-$Zn_{surf}$ and $C_{E2}$-$Zn_{surf}$) and $H^*$-abstraction. This second step has 1.15 eV activation barrier (TS2) and -1.79 eV reaction energy (FS2), calculated with respect to FS1 (**Fig. 6**). The third reaction (reaction C) is H-abstraction followed by $C_{E1}$-$O_{E1}$ bond-breaking (**ESI Fig. S5**). Our result shows the reaction is endothermic with reaction energy of ~3.0 eV. Thus, we do not explore this path further because it is energetically unfavored.

Up to this end, our results suggest that EC decomposition into $CO_2$ and $CH_2CHO$ is thermodynamically favorable in a stoichiometric ZnO surface. Further, our results also suggest that thermodynamically favorable EC decomposition reactions are always accompanied by a H-abstraction, i.e., deprotonation. This implies that EC decomposition reaction on ZnO followed the so-called proton-coupled electron transfer (PCET) mechanism. This is also consistent with previous result where EC tends to deprotonate upon contact with metal oxides, e.g., $TiO_2$ and $Nb_2O_5$ [49].

### 3.3. EC Decomposition Paths on ZnO ($10\bar{1}0$) with $O_{vac}$ and $Zn_{vac}$

The schematic of EC decomposition reactions on ZnO ($10\bar{1}0$) surface with a surface $O_{vac}$ is shown in **Fig. 7**. We designated each of the EC decomposition reaction in a similar way to what we have been used in the stoichiometric case (see **section 3.2**). Herein, in addition to $CO_2$ and $CH_2CHO$, we also included $C_2H_4$ as a possible EC decomposition product. Three EC decomposition reactions were considered: the first reaction (reaction D) produced $CO_2$, $CH_2CHO^*$, and $H^*$, while the other two reaction paths (reactions E and F) produced $CO_2/CO_2^*$, $C_2H_4$, and adsorption of $O_C$ on the $O_{vac}$ site (thus, 'healing' the vacancy site).

The first reaction (reaction D) starts from a similar EC adsorption configuration with reaction A of the stoichiometric surface. The decomposition also proceeds in the same manner as the reaction A where $H_1$-abstraction and $C_{E2}$-$O_{E2}$ ring-opening reactions are assumed to occur simultaneously (**Fig 8**). Our result shows that reaction D is an exothermic reaction with -0.59 eV reaction energy. The calculated activation barrier is 1.23 eV, slightly lower than its identical decomposition reaction occurred on ZnO stoichiometric surface, i.e., reaction A.

The second and third reactions (reactions E and F) are shown in **Fig. 9** and **ESI Fig. S6**, respectively. Both reactions have $C_{E1}$-$O_{E1}$ bond breaking that leads to ring-opening of EC but differ in the occurrence of deprotonation. Reaction E is a ring-opening reaction without dehydrogenation with $CO_2^*$ and $C_2H_4$ as the final products. As shown in **Fig. 9** the first step is the $C_{E1}$-$O_{E1}$ ring-opening followed by $O_{E1}$ adsorption on the $Zn_{surf}$. The calculated activation barrier and reaction energy for this step are 1.28 eV (TS1) and 1.16 eV (FS1), respectively. The reaction then proceeds to the second step where two simultaneous dissociations occur, i.e., the dissociation of $C_2H_4$ and $O_C$. With respect to the FS1, the second step has 0.38 eV and -2.22 eV activation barrier and reaction energy, respectively. The third reaction (reaction F) is essentially similar to the reaction E but is now accompanied by a H-abstraction (**ESI Fig. S6**). While the reaction energy is calculated to be exothermic (-0.13 eV), this reaction has a high activation barrier of 2.58 eV. This high activation barrier may kinetically prohibit the reaction to further occur.

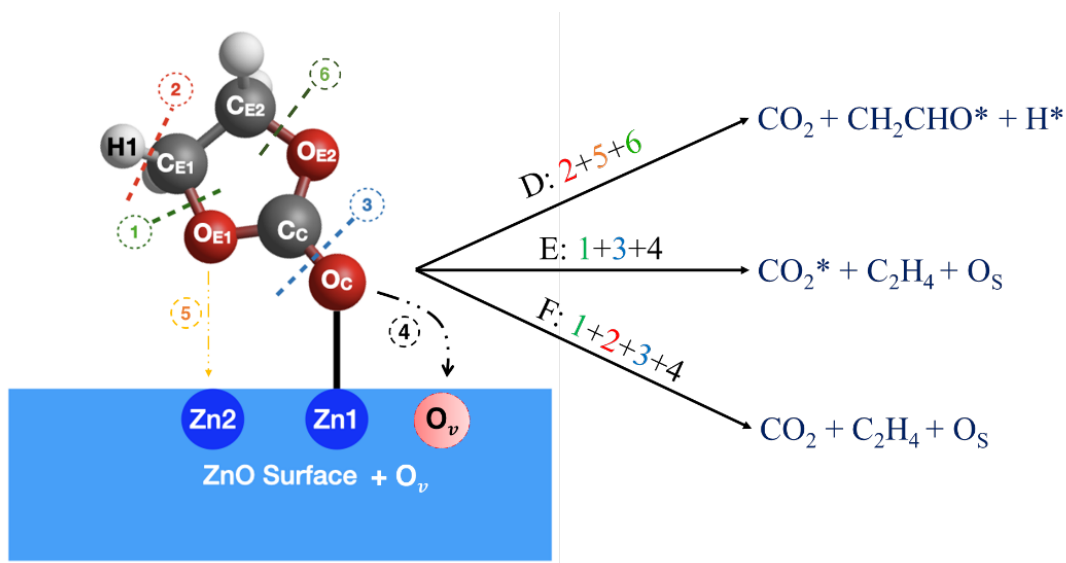

**Figure 7.** Schematic of initial decomposition reactions of EC on the ZnO ($10\bar{1}0$) surface with an O vacancy.

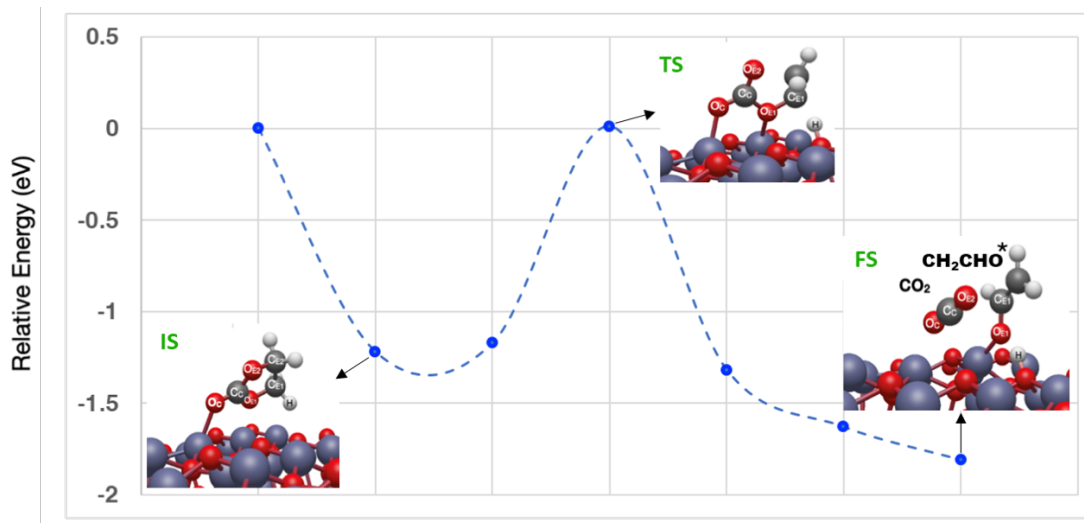

**Figure 8.** Potential energy curve for EC decomposition to $CO_2$, $CH_2CHO^*$, and $H^*$ (reaction D: 2+5+6) on the ZnO $(10\bar{1}0)$ surface with an O vacancy.

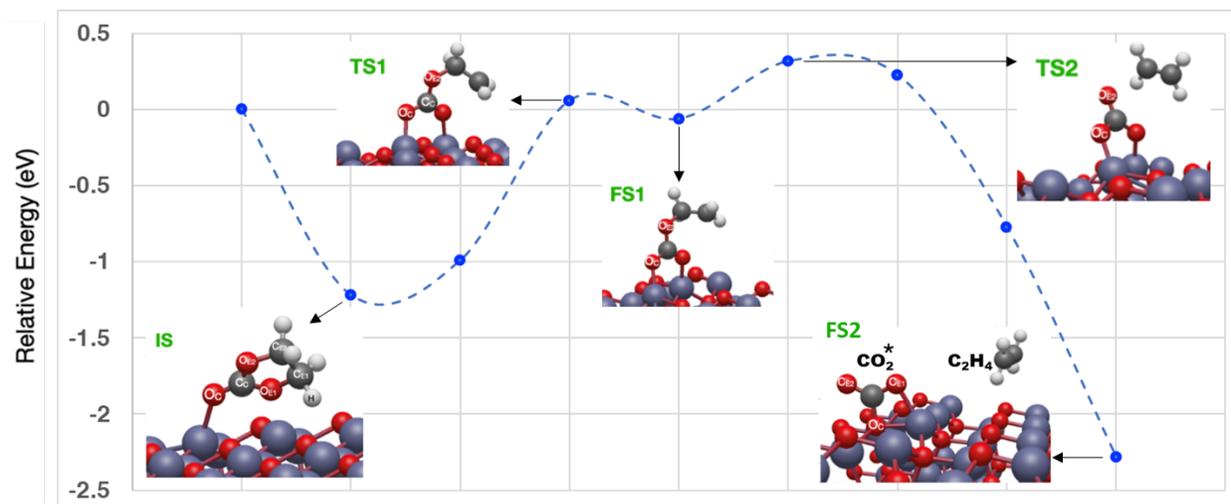

**Figure 9.** Potential energy curve for EC decomposition to $CO_2^*$ + $C_2H_4$ + $O_S$ (reaction E: 1+3+4, without H dissociation) on the ZnO $(10\bar{1}0)$ surface with an O vacancy.

Next, for the case of EC adsorption on ZnO $(10\bar{1}0)$ with a $Zn_{vac}$, we found that ring-opening reaction is unlikely to occur on the $Zn_{vac}$ site. Assuming that the ring-opening occurs via the $C_{E2}$-$O_{E2}$ bond-breaking (**Fig. 10a**), repulsive interaction between the $O_{E1}$ and the nearest two two-fold $O_{surf}$ (**Fig. 2c**) prohibits the adsorption of the ring-opened EC. Following the cases of EC decomposition on stoichiometric and $O_{vac}$-contained ZnO $(10\bar{1}0)$, we then examined the possibility of EC to reorient its adsorption configuration closer to the nearest $Zn_{surf}$ (designated with '0' in **Fig. 10a**). Calculated adsorption energies of reoriented EC in the vicinity of $Zn_{vac}$ site show almost similar value to the most stable adsorption case (**ESI Fig. S3b-d**). Based on the aforementioned result, we took the second most stable EC adsorption configuration (**ESI Fig. S3b**, $E_{ad}$ = -1.34 eV) as the first 'final state' (FS1 in **Fig. 10b**) of the reoriented EC that makes $O_E$ closer to the $Zn_{surf}$. The activation barrier for this EC reorientation is 0.53 eV. We then calculated subsequent EC decomposition with the same decomposition mechanism as reactions A and D on the stoichiometric and $O_{vac}$-contained ZnO surface. The subsequent EC decomposition reaction has a 0.96 eV activation energy (TS2 in **Fig. 10b**) and -0.47 eV reaction energy (FS2 in **Fig 10b**) calculated with respect to the FS1.

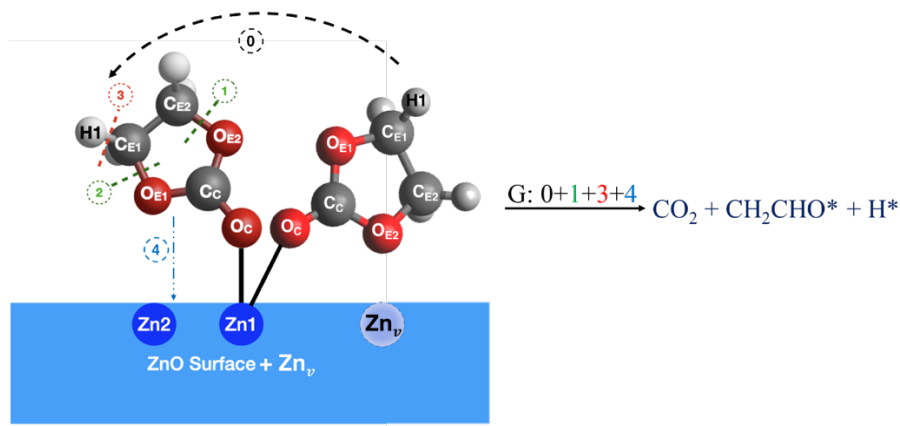

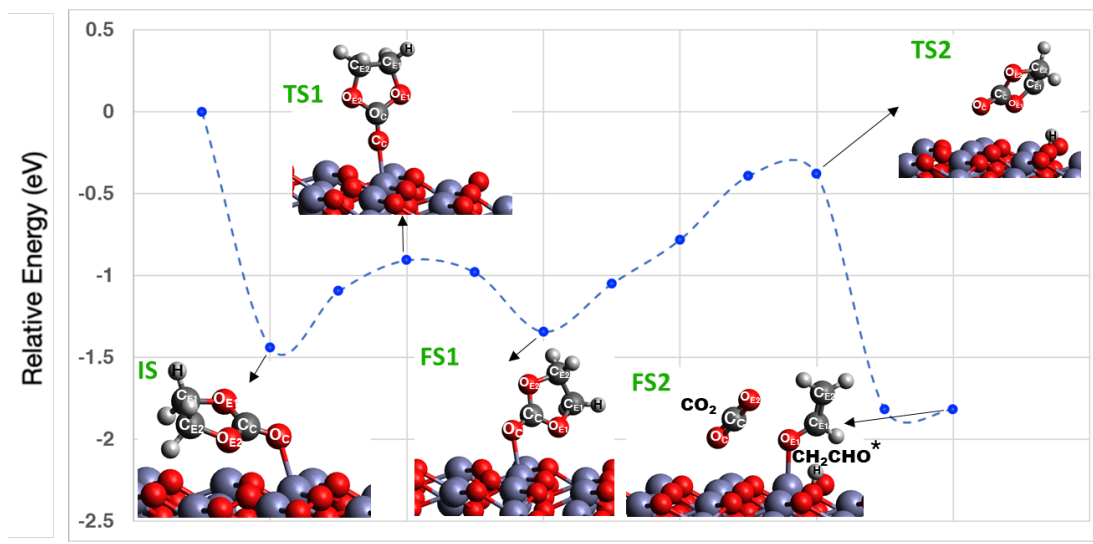

**Figure 10.** (a) Schematic of initial decomposition reaction and (b) Potential energy curve for EC decomposition to $CO_2$, $CH_2CHO^*$, and $H^*$ (reaction G: 0+1+3+4) on the ZnO $(10\bar{1}0)$ surface with a Zn vacancy.

### 3.4. Comparison of EC decomposition on ZnO and on layered cathode materials

| Surface | Reaction | Activation Barrier (eV) | Reaction Enthalpy Relative to Adsorbed State (eV) | Reaction Enthalpy Relative to Gas Phase (eV) |
|---|---|---|---|---|
| Stoichiometric | A | 1.39 | -0.58 | -1.86 |
| | B | 1.15 | -1.35 | -2.63 |
| | C | Unstable | Unstable | Unstable |
| With $O_{vac}$ | D | 1.23 | -0.59 | -1.81 |
| | E | 1.28 | -1.07 | -2.22 |
| | F | 2.58 | -0.13 | -1.35 |
| With $Zn_{vac}$ | G | 0.96 | -0.38 | -1.82 |

**Table 1.** Activation barrier and reaction energy (enthalpy) for each reaction. The activation barriers are the rate-limiting ones, while the reaction enthalpies are for each overall reaction.

We here summarized all calculated activation barriers and energies (i.e., enthalpies) in **Table 1** and also highlight several key results of EC decomposition on ZnO surface. Of the three studied reactions in stoichiometric ZnO ($10\bar{1}0$) case, reaction B has the lowest activation barrier (1.15 eV) and the largest thermodynamic driving force with reaction energy of -1.35 eV. For $O_{vac}$-contained ZnO ($10\bar{1}0$) case, reaction D has the lowest activation energy of 1.23 eV. As for reaction E, it is a two-step reaction with a rate-limiting barrier of 1.28 eV but with the most energetically favored reaction energy (-1.07 eV). For the $Zn_{vac}$-contained ZnO case, we have calculated the reaction G that is a two-step decomposition reaction with 0.96 eV rate-limiting activation barrier and an overall -0.38 eV reaction energy.

From the obtained results, we can see that both defects affect the EC decomposition activation barrier. The $O_{vac}$ ($Zn_{vac}$) slightly increase (decrease) the barrier by 0.08 (0.19 eV) relative to the lowest barrier on the pristine surface, respectively. This makes activation barriers for EC decomposition on pristine and defective ZnO ($10\bar{1}0$) surfaces range between 0.96-1.23 eV. These results imply that the presence of under-coordinated Zn-O bond (be it in the form of $O_{vac}$ and $Zn_{vac}$) on ZnO surface may kinetically increase or decrease the decomposition rate of EC compared to the pristine case. It is worth mentioning that even though $Zn_{vac}$ reduce the activation barrier, it also inhibits the ring-opening of EC through repulsive interaction with the two-fold $O_{surf}$ as previously stated.

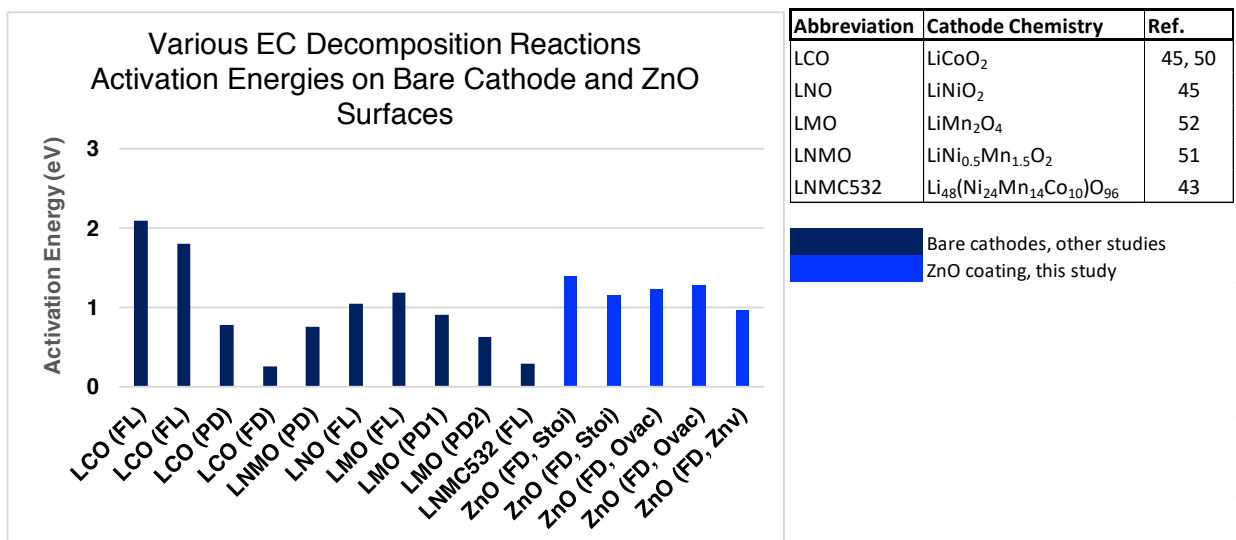

**Figure 11.** (a) EC decomposition reactions activation barriers on bare cathodes taken from references **43, 45** and **50-52** and ZnO ($10\bar{1}0$) pristine and defective surfaces. FL, PD, and FD stand for fully lithiated, partially delithiated, and fully delithiated, respectively. PD1 and PD2 stand for partially delithiated stage 1 and stage 2, respectively. 'Stoi' and $O_{vac}$ stand for stoichiometric surface and surface with an $O_{vac}$.

Next, we compare our results of the energetics of EC decomposition on both pristine and defective ZnO surface with previous results of EC decomposition on several layered oxide cathodes. For this purpose, we first tabulated and compared the activation energy needed to break the $C_E$-$O_E$ bond, i.e., the ring-opening reaction of EC (**Fig. 11**). We would like to remind again that all the reactions studied here were assumed to be chemical in nature. This means that any external potential was excluded. Therefore, the data that we have taken to be compared with our results were also calculated with similar condition to ours in order to have a justifiable comparison (i.e., modeled in ultra-high vacuum and without external potential). We can see from **Fig. 11** that the ring-opening reaction of EC on ZnO is more difficult to occur than on most of the LIB cathodes (viz., LNO (FL), LMO (FL), LCO (PD), LNMO (PD), LCO (FD)) [43,45,50–52]. Only fully-lithiated LCO that has higher propensity against EC decomposition as compared to the ZnO. However, it should be noted that once the LiCoO$_2$ is in a partially delithiated state, the activation barrier of the ring-opening reaction is significantly decrease by about 1.0 eV [45]. Further, fully-delithiated LCO (LCO (FD)) shows a fast ring-opening reaction of EC with only 0.25 eV activation barrier [45]. This comparison shows that in general, even with the presence of defects, ZnO has the better propensity against surface-mediated catalytic EC decomposition reaction than bare LIB cathode. Our finding is consistent with previous results that showed ZnO-coated cathodes of LIB always have a better capacity retention than LIB with untreated cathode.

We also assess our calculated activation barrier of EC decomposition on ZnO with regard to the possible working condition of a LIB. Firstly, we compare the activation barrier of EC on ZnO with an expected activation barrier needed to hinder EC decomposition in an idealized LIB environment. By assuming an Arrhenius reaction rate with prefactor of $k_bT/h$, Xu *et al* calculated that at least 1.03 eV of a rate-limiting activation barrier is needed to hinder EC decomposition [43]. This value was calculated by taking into account the following assumptions: (1) the working condition of the LIB follows the hybrid electric vehicle (HEV) working condition where at most 10% of the electrolyte are decomposed within one year, (2) the LIB working temperature can be kept constant at T=300 K (within the one year working condition) and (3) the mass percentage of the HEV LIB consists of 27% and 4.4% of cathode and electrolyte, respectively. Comparing the value calculated by Xu *et al* with our calculated EC decomposition activation barriers, it can be inferred that in general ZnO fulfill the required minimum activation barrier. Secondly, note that a protective coating material for LIB cathode should also facilitate facile Li diffusion. Though it is beyond the scope of the present work, we would like to point out briefly how the two considered point defects here might affect Li diffusion inside ZnO. Previous studies have shown that $Zn_{vac}$ inside ZnO tends to trap interstitial Li ($Li_i$) subsequently forming the $Li_{Zn}$ antisite defect ($Li_i$ + $Zn_{vac}$ → $Li_{Zn}$) [53]. The $Li_{Zn}$ defect was found to be stable with reverse reaction (or dissociation energy) for $Li_{Zn}$ → $Li_i$ + $Zn_{vac}$ to be endoenthalpic with 3.57 eV. Intuitively, this result suggests that the presence of $Zn_{vac}$ may hinder the Li mobility inside the ZnO. For the case $O_{vac}$ containing ZnO, several studies have shown that the presence of $O_{vac}$ can increase the rate of Li diffusion inside the ZnO [28,54]. Considering that EC decomposition can still be suppressed on $O_{vac}$-containing ZnO, synthesizing high-energy cathode coated with O deficient ZnO can be a promising strategy to enhance capacity retention, electronic conductivity and ionic diffusivity at the same time. However, careful consideration of the $O_{vac}$ amount should also be taken into account. This is because too high $O_{vac}$ concentration can also facilitate O migration from the cathode that may trigger other deleterious effect such as phase transformation of the active cathode [28].

## 4. Summary

We have studied ethylene carbonate adsorption and decomposition reactions on stoichiometric and defective ZnO (10$\bar{1}$0) surface by means of first-principles density functional theory (DFT) calculations. Here, we used $CO_2$, $C_2H_4$, $CH_2CHO$ and H as probing compounds to evaluate the thermodynamic and kinetic of EC decomposition mediated by pristine and defective ZnO (10$\bar{1}$0). Our results reveal that in the vicinity of $O_{vac}$, activation barriers of EC decomposition are higher compared to the decomposition on pristine ZnO (10$\bar{1}$0), while the reverse is true for $Zn_{vac}$. However, the $Zn_{vac}$-contained ZnO (10$\bar{1}$0) surface has lesser active sites due to repulsion between $O_{E1}$ and the nearest two-fold $O_{surf}$. Comparison of EC decomposition activation barriers on several LIB cathodes with our results suggests that ZnO has an overall better propensity against EC decomposition than LIB bare cathodes. Furthermore, the lowest activation barrier found for EC on $Zn_{vac}$-contained ZnO (10$\bar{1}$0) shows a value (0.96 eV) that is close to the minimum required activation barrier to hinder rapid EC decomposition in a LIB working environment (1.03 eV). Our findings suggest that ZnO is a good material for LIB cathode protection, in particular against chemically driven EC decomposition. Finally, our present study can be used as a guideline to further evaluate liquid electrolyte interaction with other candidates of protective coating materials for LIB cathode. We propose that finding a coating material that has good thermodynamic and kinetic stability against the liquid carbonate solvent is also important aside from other criteria such as stability against salt decomposition.

**Acknowledgement**

A.G.S. would like to acknowledge research fund granted by the Directorate of Higher Education, Ministry of Research Technology and Higher Education (RISTEKDIKTI), Republic of Indonesia under grant scheme "Penelitian Dasar Unggulan Perguruan Tinggi 2021". H.K.D. acknowledged research fund granted by the "Indonesia-MIT Program for Advanced Research and Technology" (IMPACT). All calculations were performed using the High-Performance Computing facility at the Research Center for Nanosciences and Nanotechnology (RCNN), Institut Teknologi Bandung.

**Corresponding Authors**

*Adhitya Gandaryus Saputro: ganda@tf.itb.ac.id

*Ganes Shukri: ganes_tf07@itb.ac.id